\documentclass[11pt]{article}
\usepackage{amssymb}\usepackage{amsmath}
\usepackage[all]{xy}
\usepackage[dvips]{graphicx}
\numberwithin{equation}{section}

\def\be{\begin{equation}}
\def\ee{\end{equation}}
\def\ba{\begin{align}}
\def\ea{\end{align}}
\def\beq{\begin{eqnarray}}
\def\eeq{\end{eqnarray}}

\def\yboxit#1#2{\vbox{\hrule height #1 \hbox{\vrule width #1
\vbox{#2}\vrule width #1 }\hrule height #1 }}
\def\fillbox#1{\hbox to #1{\vbox to #1{\vfil}\hfil}}
\def\ybox{{\lower 1.3pt \yboxit{0.4pt}{\fillbox{8pt}}\hskip-0.2pt}}
\def\comments#1{}

\def\II{\relax{I\kern-.10em I}}

\def\IZ{\relax\ifmmode\mathchoice
{\hbox{\cmss Z\kern-.4em Z}}{\hbox{\cmss Z\kern-.4em Z}}
{\lower.9pt\hbox{\cmsss Z\kern-.4em Z}}
{\lower1.2pt\hbox{\cmsss Z\kern-.4em Z}}\else{\cmss Z\kern-.4em
Z}\fi}
\def\IB{\relax{\rm I\kern-.18em B}}
\def\IC{{\relax\hbox{$\inbar\kern-.3em{\rm C}$}}}
\def\ID{\relax{\rm I\kern-.18em D}}
\def\IE{\relax{\rm I\kern-.18em E}}
\def\IF{\relax{\rm I\kern-.18em F}}
\def\IG{\relax\hbox{$\inbar\kern-.3em{\rm G}$}}
\def\IGa{\relax\hbox{${\rm I}\kern-.18em\Gamma$}}
\def\IH{\relax{\rm I\kern-.18em H}}
\def\II{\relax{\rm I\kern-.18em I}}
\def\IK{\relax{\rm I\kern-.18em K}}
\def\IN{\relax{\rm I\kern-.18em N}}
\def\IP{\relax{\rm I\kern-.18em P}}

\def\inbar{\,\vrule height1.5ex width.4pt depth0pt}

\font\cmss=cmss10 \font\cmsss=cmss10 at 7pt
\def\IR{\relax{\rm I\kern-.18em R}}

\def\lp10{l_P^{10}}
\def\lp11{l_P^{11}}
\def\R11{R_{11}}

\font\manual=manfnt
\def\dbend{\lower3.5pt\hbox{\manual\char127}}

\def\IZ{\relax\ifmmode\mathchoice
{\hbox{\cmss Z\kern-.4em Z}}{\hbox{\cmss Z\kern-.4em Z}}
{\lower.9pt\hbox{\cmsss Z\kern-.4em Z}} {\lower1.2pt\hbox{\cmsss
Z\kern-.4em Z}}\else{\cmss Z\kern-.4em Z}\fi}

\def\rt2{\sqrt{2}}
\def\irt2{\frac{1}{\sqrt{2}}}

\def\a{\alpha}

\font\cmss=cmss10
\font\cmsss=cmss10 at 7pt
\def\IL{\relax{\rm I\kern-.18em L}}
\def\IH{\relax{\rm I\kern-.18em H}}
\def\IR{\relax{\rm I\kern-.18em R}}
\def\inbar{\vrule height1.5ex width.4pt depth0pt}
\def\IC{\relax\hbox{$\inbar\kern-.3em{\rm C}$}}
\def\rlx{\relax\leavevmode}
\def\ZZ{\rlx\leavevmode\ifmmode\mathchoice{\hbox{\cmss Z\kern-.4em Z}}
 {\hbox{\cmss Z\kern-.4em Z}}{\lower.9pt\hbox{\cmsss Z\kern-.36em Z}}
 {\lower1.2pt\hbox{\cmsss Z\kern-.36em Z}}\else{\cmss Z\kern-.4em
 Z}\fi}
\def\IZ{\relax\ifmmode\mathchoice
{\hbox{\cmss Z\kern-.4em Z}}{\hbox{\cmss Z\kern-.4em Z}}
{\lower.9pt\hbox{\cmsss Z\kern-.4em Z}}
{\lower1.2pt\hbox{\cmsss Z\kern-.4em Z}}\else{\cmss Z\kern-.4em
Z}\fi}

\font\manual=manfnt
\def\dbend{\lower3.5pt\hbox{\manual\char127}}

\def\IZ{\relax\ifmmode\mathchoice
{\hbox{\cmss Z\kern-.4em Z}}{\hbox{\cmss Z\kern-.4em Z}}
{\lower.9pt\hbox{\cmsss Z\kern-.4em Z}} {\lower1.2pt\hbox{\cmsss
Z\kern-.4em Z}}\else{\cmss Z\kern-.4em Z}\fi}

\def\rt2{\sqrt{2}}
\def\irt2{\frac{1}{\sqrt{2}}}

\title{\Large{\bf On the Classification of Residues of the Grassmannian}} 
\author{Sujay K. Ashok$^{a,b}$ and Eleonora Dell'Aquila$^{b}$ } \date{}
\begin{document}
\maketitle

\begin{center}
  $^{a}$Institute of Mathematical Sciences\\
  C.I.T Campus, Taramani\\
  Chennai, India 600113\\
  \vspace{.3cm}
  $^{b}$Perimeter Institute for Theoretical Physics\\
  Waterloo, Ontario, ON N$2$L$2$Y$5$, Canada \\
\end{center}

 \begin{abstract}
 We study leading singularities of scattering amplitudes which are obtained as residues of an integral over a  Grassmannian manifold. 
 We recursively do the transformation from twistors to momentum twistors and obtain an iterative formula for Yangian invariants that involves a succession of dualized twistor variables. 
 This turns out to be useful in addressing the problem of classifying the residues of the Grassmannian. 
 The iterative formula leads naturally to new coordinates on the Grassmannian in terms of which both composite and non-composite residues appear on an equal footing. 
 We write down residue theorems in these new variables and classify the independent residues for some simple examples. 
 These variables also explicitly exhibit the distinct solutions one expects to find for a given set of vanishing minors from Schubert calculus. 
\end{abstract}

\newpage

\tableofcontents


\section{Introduction}

A major breakthrough in the study of scattering amplitudes of ${\cal N}=4$ SYM was the conjecture made in \cite{ACCK0907}, that leading singularities of the N$^{k}$MHV amplitudes involving $n$ particles can be obtained as residues of the following integral: 
\begin{multline}\label{Grassmannian}
{\cal L}_{n,k+2}(\lambda,\tilde\lambda, \tilde\eta) = \frac{1}{\text{Vol} (GL(k+2))}\ 
\int \frac{d^{(k+2)\times n}C_{\alpha a}}{C(12\ldots k+2)\ldots C(n1\ldots k+1)}\cr 
\times \prod_{\alpha=1}^{k+2}\delta^{2}(C_{\alpha a}\tilde\lambda_a)\delta^4(C_{\alpha a}\tilde \eta_a)
\prod_{\beta=1}^{n-k-2}\delta^{2}(\tilde{C}_{\beta a}\lambda_a) \ .
\end{multline}
Here $(\lambda_a, \tilde\lambda_a)$, with $a\in \{1, \ldots. n\}$ label the null external momenta of the scattering particles and, geometrically, they specify two distinct $2$-planes in $\mathbb{C}^n$. The $\tilde\eta_i$ are Grassmann variables which keep track of the helicity of the scattering particles.  
The delta functions in \eqref{Grassmannian} constrain the $2$-plane $\tilde{\lambda}$ to be orthogonal to a $k+2$-plane described by the matrix $C\,$, and $\lambda$ to be orthogonal to the complementary $n-k-2$-plane, denoted by $\tilde{C}$. This ensures that $\lambda$ and $\tilde{\lambda}$ are orthogonal planes, which is equivalent to demanding momentum conservation. 
The measure of the integral is the product of $n$ factors $C(i,i+1,\ldots, i+k+2)$, each a minor computed from $(k+2)$ consecutive columns, and it ensures that the integral is invariant under a $GL(k+2)$ transformation on $C$. Therefore the integral in \eqref{Grassmannian} is defined over the space of $k+2$-planes in $\mathbb{C}^n\,$, which is the Grassmannian $G(n, k+2)$ \cite{ACCK0907}. 

The Grassmannian integral in \eqref{Grassmannian} was first presented in twistor space variables \cite{Penrose:1967wn, Penrose:1968me, Penrose:1972ia, Penrose:1999cw} as a way to encode the properties of scattering amplitudes and the BCFW recursion relations in twistor space \cite{Witten:2003nn}-\cite{Korchemsky:2009jv}. Moreover, it was also shown that it reflects the integrable structure of ${\cal N}=4$ SYM and possesses all the expected symmetries \cite{ACCK0907, MS0909, ACC0909}. As is well known, the planar limit of the ${\cal N}=4$ theory has a Yangian symmetry \cite{DNW0308, DNW0401}. The Yangian has a set of generators that correspond to the usual superconformal generators and another set that generates a dual superconformal symmetry \cite{Alday:2007hr}-\cite{McGreevy:2008zy}. These are, respectively, the level zero and level one generators of the Yangian algebra (in one presentation). The Grassmannian integral has been shown to be invariant under this full Yangian symmetry \cite{DF1001}. Furthermore, it is argued in \cite{DF1002, KS1002} that all the Yangian invariants can be generated as contour integrals in $G(k,n)$. 

The superconformal invariance of the leading singularities is best seen by writing them out in twistor space, which is done by Fourier transforming with respect to the $\lambda$ variable in \eqref{Grassmannian}, resulting in \cite{ACCK0907}:
\be\label{grassmanniantwistor}
{\cal L}_{n,k+2}({\cal W})= \frac{1}{\text{Vol} (GL(k+2))}\times\int \frac{d^{(k+2)\times n}C_{\alpha a}}{C(12\ldots k+2)\ldots C(n1\ldots k+1)}\, \prod_{\alpha=1}^{k+2}\delta^{4|4}(C_{\alpha a}{\cal W}_a)\,,
\ee 
where ${\cal W}_a = (\tilde \mu_a, \tilde\lambda_a | \tilde\eta_a)\in \mathbb{CP}^{3|4}$, which is super-twistor space. On the other hand, the dual superconformal symmetry is best seen by changing variables in \eqref{Grassmannian} to what are called momentum twistors \cite{Hodges0905}, which we denote by ${\cal Z}\,$:
\be\label{intromomtwist}
{\cal L}_{n,k+2}(\lambda, \tilde\lambda) = A_{MHV}^{\text{tree}}\, {\cal R}_{n,k}({\cal Z}) \,.
\ee
Unlike the transformation between the twistor variables ${\cal W}$ and the momenta --which is a Fourier transformation-- the  relation between the momentum twistors and momenta is a purely algebraic one. The factor ${\cal R}_{n,k}({\cal Z})$  is a Yangian invariant \cite{ABC1008} and will be the focus of our analysis in later sections. 

One interesting aspect of these manipulations is the self-similar nature of the transformations. ${\cal R}_{n,k}({\cal Z})$ is written as 
\be\label{Rnk}
{\cal R}_{n,k}({\cal Z}) = \frac{1}{\text{Vol} (GL(k))}\times\int \frac{d^{k\times n}D_{\alpha a}}{D(12\ldots k)\ldots D(n1\ldots k)}\, \prod_{\alpha=1}^{k}\delta^{4|4}(D_{\alpha a}{\cal Z}_a)\,,
\ee
which has the same form as ${\cal L}_{n,k+2}({\cal W})$ in twistor space, but with $k$ reduced by two. We will use ${\cal Y}_{n,k}({\cal Z})$ to denote the integral \eqref{Rnk} in later sections, where the contour does not necessarily pick out a leading singularity. 
A natural question to ask is whether it is possible to iterate this procedure, thereby systematically reducing the value of $k$, and define a sequence of dualized twistor variables in the process. This is what we will do in the later sections, with some modifications with respect to the original method, resulting in an iterative formula for the Yangian invariants. The formula turns out to be a useful  bookkeeping device for the classification of residues. 

Leading singularities have been conjectured to be sufficient data to reconstruct the perturbative S-matrix of ${\cal N}=4$ SYM \cite{C0808, ACK0808} and this has been checked for all one-loop amplitudes and for some examples at higher loops\footnote{See \cite{Britto} for a recent review on loop amplitudes.} \cite{Bern:1993mq}-\cite{BDKRSVV}. Since it has been shown that all leading singularities can be obtained as residues of the Grassmannian integral \eqref{Grassmannian}, we have a physical motivation to obtain a complete classification of these residues. 
 
Another motivation for the classification arises as follows: in \cite{BMS0912} it was shown that for a given $n$ and $k$, there are only certain primitive configurations in twistor space where any leading singularity can be supported. The simplest leading singularities with such a support are called primitive. This means that for a given $n$ and $k$, and to all orders in perturbation theory, there is only a finite number of different algebraic functions that a leading singularity can be. As we mentioned earlier,  leading singularities have been shown to be invariant under the Yangian symmetry \cite{ABC1008}. Moreover, it was shown that all Yangian invariants are residues of the Grassmannian. In this spirit, if it were possible to count residues and primitive singularities independently, given that the two sets for a given $n$ and $k$ are finite, one could prove their complete equivalence.

Let us now discuss in some detail how to define residues of the integral in \eqref{Grassmannian}. The description of the Grassmannian through the matrix $C$ contains a $GL(k+2)$ redundancy (or gauge symmetry), that can be eliminated by imposing $(k+2)^2$ conditions on the $C_{\alpha a}$'s (fixing the gauge). We also have $2n$ explicit delta-functions, but we know that four of these must encode momentum conservation. Therefore, there are only $2n-4$ effective delta functions that constrain the elements $C_{\alpha a}\,$. Taking all this into account, we see that the integral in  \eqref{Grassmannian} localizes onto a submanifold of dimension
\be\label{d}
d\equiv n(k+2)-(k+2)^2-(2n-4)=k(n-k-4) \,.
\ee
This means that every residue of \eqref{Grassmannian} corresponds to a different way of imposing $d$ conditions on the minors of $C$. Here the following technical difficulty arises. For generic $k\,$, the number of conditions to be imposed exceeds the number $n$ of factors that appear explicitly in the denominator.  Therefore it must be that on the locus where a given minor vanishes, another minor further factorizes, and so on. This phenomenon gives rise to what are called composite residues and a complete classification of these residues has so far not been carried out although there has been much progress along this direction \cite{ BMS0912, K0912, ABCT0912, B1008}.

We will show that, using the iterative technique mentioned above, it is possible to define new coordinates on the Grassmannian, simply defined in terms of the minors, in terms of which both composite and non-composite residues appear on an equal footing. In this work we focus on small values of $n$ where we are able to explicitly count the number of independent residues of the Grassmannian for $k=2$. All other residues can be obtained from this given set by using simple residue theorems in the new variables. 

The new variables also carry some unexpected bonuses. In \cite{ACCK0907} it was shown that the number of solutions one expects by setting the $d$ minors to zero is related to the calculation of a certain Littlewood-Richardson coefficient, which followed from calculating the self-intersection of a Schubert cycle, and this was checked for small values of $k$ and $n$. We show that in terms of the new variables it is possible to exhibit these different solutions explicitly, at least for all values of $n\le 12$ (when the number of consecutive minors is greater than or equal to $d$). 

This paper is organized as follows. In section 2, we derive our main formula, that iteratively relates the leading singularities for a given $n$ and $k$ to one with smaller $k$. The analysis parallels closely the transformation from twistor variables to momentum twistors. From then on, we will restrict our attention to the case $k=2$. In section 3, we find a general map (for all $n$) that relates vanishing minors in the original variables to specific configurations of the new ones. We will apply this map in sections 4 and 5 to map out the set of independent residues for the cases $8\le n\le 12$. We conclude with a summary of results and open questions in section 6. 

\section{Maximal breakdown of the N$^{k}$MHV residues}

\subsection{Review of Momentum Twistors}

Consider the Grasmannian integral \eqref{Grassmannian}. In the first part of this section, we will review the analysis of \cite{ACC0909} and recall how momentum twistors arise as a change of variables. 

The key is to choose a gauge such that the first two rows of the $C$ matrix coincide with $\lambda\,$. To achieve this, we introduce $k+2$ auxiliary spinor variables $\rho_{\alpha}\,$, as follows:
\begin{multline}\label{momtwistrho}
{\cal L}_{n,k+2}=\int \frac{d^{k\times n}C\, \Delta_{GL(k+2)}}{C(12\ldots k+2)\ldots C(n1\ldots k+1)}\cr
\times \prod_{\alpha=1}^{k+2}\delta^{2|4}(C_{\alpha a}\tilde\lambda_a) \int \prod_{\a=1}^{k+2}d^2\rho_{\alpha} \prod_{a=1}^{n}\delta^{2}(\lambda_a - C_{\alpha a}\rho_{\alpha})\,.
\end{multline}
Here the $GL(k+2)$ redundancy is lifted by explicitly including the $\Delta_{GL(k+2)}$ factor, which should be thought of as a product of $(k+2)^2$ delta functions that fix some components of $C$ and $\rho\,$. 
Out of these, $2(k+2)$ delta functions can be used to set
\be\label{rhofixed}
\rho^{T} = \left(\begin{array}{ccccc}
1 & 0 & 0 & \ldots & 0 \cr
0 & 1 & 0 & \ldots & 0 
\end{array} \right) \,
\ee
and as a result, $C$ takes the form
\be\label{Clambda}
C = \left(\begin{array}{ccccc}
\lambda_{1,1} &\lambda_{1,2} & \ldots & \lambda_{1,n} \cr
\lambda_{2,1} &\lambda_{2,2} & \ldots & \lambda_{2,n}\cr
C_{3,1} & C_{3,2} & \ldots &C_{3,n} \cr
\vdots & \vdots & \ddots & \vdots \cr
C_{k+2,1} & C_{k+2,2} & \ldots & C_{k+2,n} \end{array}\right)\,.
\ee

In order to reduce $\Delta_{GL(k+2)}$ to $\Delta_{GL(k)}\,$, we still need to impose $2k$ delta functions. 
This reflects the fact that the condition \eqref{rhofixed} leaves a residual symmetry, because we can still translate the $k$ remaining rows of $C$ in  \eqref{Clambda} by either $\lambda_{1,a}$ or $\lambda_{2,a}$ without changing the minors of $C$. This can be taken care of by introducing
\be
J(\lambda)\, \prod_{\hat\alpha=3}^{k+2}\delta(C_{\hat\alpha a} \lambda_a) \,,
\ee
where $J$ is the resulting Jacobian factor that depends only on $\lambda$. Putting all this together, we get
\begin{multline}
{\cal L}_{n,k+2} = J(\lambda)\delta^4(\lambda_a\tilde\lambda_a)\delta^8(\lambda_a\eta_a)\int \frac{d^{k\times n}C_{\hat\alpha a}\ \Delta_{GL(k)}}{C(12\ldots k+2)\ldots C(n1\ldots k+1)} \cr
\times \prod_{\hat\alpha=3}^{k+2} \delta^2(C_{\hat\alpha a}\lambda)\, \delta^{2|4}(C_{\hat\alpha a}\tilde \lambda)\,.
\end{multline}

The next step is to change variables to some $k\times n$ matrix $D$ such that the minors of $C$ have a simple expression in terms of the minors of $D$. This can be implemented by inserting the identity as
\be
1 = \int d^{k\times n}D_{\hat\alpha a} \prod_{\hat\alpha, a} \delta(D_{\hat\alpha a} - C_{\hat\alpha a} Q_{ab}) \,,
\ee
where
\begin{align}
Q_{ab} &= \frac{\langle\lambda_{b+1},\lambda_b\rangle\, \delta_{a,b-1}+\langle\lambda_{b-1}, \lambda_{b+1}\rangle\, \delta_{a,b} + \langle\lambda_b, \lambda_{b-1}\rangle\, \delta_{a, b+1}}{\langle\lambda_{b-1}, \lambda_b\rangle\langle\lambda_{b}, \lambda_{b+1}\rangle} \,.
\end{align}
Then we change variables to $\tilde\lambda_a = Q_{ab}\mu_b\,$ (and corresponding fermionic variables  $\tilde\eta$'s), and integrate over the elements of $C$ to be left with an integral over the  elements of $D$ of the form 
\begin{align}\label{newgrass}
{\cal L}_{n,k+2} &= \frac{A_{n,MHV}^{\tt tree}(\lambda,\tilde\lambda)}{\text{Vol}(GL(k))}\int \frac{d^{k\times n}D_{\alpha a}}{D(12\ldots k)\ldots D(n1\ldots k-1)} \prod_{\alpha=1}^{k} \delta^{4|4}(D_{\alpha a}{\cal Z}^m_a) \cr
&= A_{n,MHV}^{\tt tree}(\lambda,\tilde\lambda)\, {\cal R}_{n,k}({\cal Z})\,.
\end{align}
Here ${\cal Z}^m_a = (\lambda_a, \mu_a, \eta_a)$ are the momentum twistors \cite{Hodges0905, MS0909, ACC0909} and one can see that the integral is now defined over the Grassmannian $G(n,k)$, parametrized by the $k\times n$ matrix $D$. In the derivation, the Jacobian $J(\lambda)$ is fixed by requiring the expression to transform appropriately under the little group transformations. We refer the reader to \cite{ACC0909} for details. 

\subsection{New variables from a $3+1$ split} 

We would like to iterate the procedure that led to \eqref{newgrass} from \eqref{Grassmannian} by starting with the expression for ${\cal R}_{n,k}({\cal Z})$ and attempt to reduce the value of $k$. However, we will find it more convenient at this stage to do a $3+1$ split of the momentum-twistor $Z^m_a= (\tilde Z_a, z_a)$, where $\tilde Z_a$ is a 3-vector and $z_a$ is a single component. We will suppress the supersymmetric indices in what follows and reinstate them in the final answer in order to avoid cluttering the formulae. 

As mentioned in the introduction, when we do not specify a specific contour to calculate the leading singularity we will refer to the integral expression in \eqref{Rnk} as ${\cal Y}_{n,k}$: 
\begin{align}\label{oldRnk}
{\cal Y}_{n,k}({\cal Z}) &= \frac{1}{\text{Vol}(GL(k))}\int \frac{d^{k\times n}D_{\alpha a}}{D(12\ldots k)\ldots D(n1\ldots k-1)} \prod_{\alpha=1}^{k} \delta^{3}(D_{\alpha a}\tilde Z_a)\delta(D_{\alpha a}z_a) 
\end{align}
and rewrite it as 
\begin{multline}\label{oldRnkFT}
{\cal Y}_{n,k}({\cal Z}) =  \frac{1}{\text{Vol}(GL(k))}\ \times\\ \times
\int \frac{d^{k\times n}D_{\alpha a}}{D(12\ldots k)\ldots D(n1\ldots k-1)} \prod_{\alpha=1}^{k} \delta^{3}(D_{\alpha a}\tilde Z_a) \int \prod_{a}dw_a e^{iw_az_a} \int d\rho_{\alpha} \prod_{a=1}^n\delta(w_a-\rho_{\alpha}D_{\alpha a})\,.
\end{multline}
Note that, apart from the Fourier transform, this expression is very similar to the one we obtained in \eqref{momtwistrho}. From now on, our manipulations will closely parallel those in the previous section with $w_a$ playing the role of $\lambda_a$. 

Following the earlier analysis, we would like to break the $GL(k)$ gauge symmetry to its $GL(k-1)$ subgroup in an attempt to effectively reduce the rank of the matrix $D$ by one unit. Out of the $2k-1$ parameters that we need to fix, $k$ of them will be used to set $\rho_{\alpha}$ to be of the form $(1,0,\ldots 0)\,$. The $n$ delta functions $\prod_{a=1}^{n}\delta(w_a-\rho_\alpha D_{\alpha a})$ force the first row of $D$ to be identified with the vector $w_a\,$:
\be
D = \left(\begin{array}{ccccc}
w_{1} &w_{2} & \ldots &w_{n} \cr
D_{2,1} & D_{2,2} & \ldots &D_{2,n} \cr
\vdots & \vdots & \ddots & \vdots \cr
D_{k,1} & D_{k,2} & \ldots & D_{k,n} \end{array}\right)\,.
\ee
We still have the freedom to translate the remaining $k-1$ rows of $D_{\alpha a}$ by $w_a\,$. This can be fixed by imposing that each row must be orthogonal to $w_a\,$:
\be\label{gaugefix}
\sum_{a=1}^{n}D_{\hat\alpha a}w_a=0\quad\text{for}\quad \hat\alpha=2,\ldots, k \,.
\ee
This gives us the remaining $k-1$ constraints we need to break $GL(k)$ to $GL(k-1)$. Denoting the corresponding Jacobian by $J'(w)\,$, we get  
\begin{multline}
{\cal Y}_{n,k}=\int dw_a e^{iw_az_a}\delta^{3}(\tilde{Z}_a w_a)\frac{J'(w)}{\text{Vol}(GL(k-1))}\ \times \\
\times\int \frac{d^{(k-1)\times n}D_{\hat\alpha a}}{D(12\ldots k)\ldots D(n1\ldots k-1)}\prod_{\hat\alpha =2}^{k}\delta^3(D_{\hat\alpha a} \tilde Z_a)\delta(D_{\hat\alpha a}w_a) \,.
\end{multline}
We now have to express the $k\times k$ minors of $D$ in terms of $w_a$ and the remaining $k-1$ rows of $D$. In order to facilitate this we insert, as before, an identity operator as the integral over an auxiliary $t$-matrix:
\be\label{identity}
1 = \int \prod_{b=1}^n \prod_{\hat\alpha=2}^{k} dt_{\hat \alpha b} \delta(t_{\hat\alpha b} - D_{\hat\alpha a}Q_{ab}) \,,
\ee
where $Q_{ab}$ is an $n\times n$ matrix which satisfies $w_a Q_{ab}=0\,$. There are many possible choices for $Q_{ab}$; we will choose
\be
Q_{ab} = \frac{1}{w_{b}^2 w_{b+1}}(\delta_{a,b}w_{b+1}-\delta_{a, b+1}w_b) \,.
\ee
With this choice, the $k\times k$ minors of $D_{\alpha a}$ are simply proportional to the $(k-1)\times (k-1)$ minors of $t_{\hat\alpha a}$, with a proportionality constant that is a function of $w$:
\be
D(1,2,\ldots, k) =  f(w)t(1,2,\ldots, k-1) \,.
\ee
Note that unlike the case discussed in \cite{ACC0909},  $Q_{ab}$  is not symmetric in our case. However, notice that the $t$'s still satisfy the $(k-1)$ constraints
\be
\sum_{b=1}^{n}w_b t_{\hat\alpha b} = 0 \,.
\ee
It follows that in order to completely integrate over $D$, we cannot simply use the $n(k-1)$ delta functions that involve the variables $t$ since there are $k-1$ relations among them. Fortunately, there are precisely $k-1$ delta functions involving the $D_{\alpha a}$ that arose out of gauge fixing the translation symmetry along $w_a$. These two sets of delta functions can therefore be traded off for each other and the integral over $D$ performed, leading to an overall Jacobian factor that depends only on the $w_a$'s. The expression for ${\cal Y}_{n,k}$ can therefore be simplified to the form
\begin{multline}
{\cal Y}_{n,k}= \int dw_a e^{iw_az_a}\delta^{3}(\tilde{Z}_a w_a)\frac{J(w)}{\text{Vol}(GL(k-1))} \ \times\\
\times\int \frac{d^{(k-1)n}t_{\alpha a}}{t(1,2,\ldots k-1)\ldots t(n,1,\ldots k-2)} \prod_{\alpha=1}^{k-1}\delta^4(t_{\alpha a}Z^D_{a}) \,.
\end{multline}
Here we have relabelled $\hat\alpha$ back to $\alpha$ and we have shifted the range so that $\alpha$  runs from $1$ to $k-1\,$. We have also introduced new ``dual" momentum-twistors, defined so that 
\be
Z^D_a = (w_a, \mu_a) \quad \text{where}\quad \tilde Z_a = Q_{ab}\mu_b \,,
\ee
in analogy with the usual momentum twistors. The overall Jacobian $J(w)$ can be fixed by requiring that the function ${\cal Y}_{n,k}$ have the correct little group transformations under \mbox{$Z^m_a\to \xi_a\, Z^m_a\,$}. Reintroducing the fermionic delta functions, we obtain the final (supersymmetric) formula
\begin{align}\label{recursion}
{\cal Y}_{n,k}({\cal Z}) &= \int \prod_{a=1}^n \frac{dw_a}{w_a} e^{iw_az_a} \delta^{3|4}(w_a\tilde Z_a)\frac{1}{\text{Vol}(GL(k-1))} \times\cr &\qquad\qquad\qquad\times\int  \frac{dt_{\alpha a}}{t(1,2,\ldots k-1)\ldots t(n,2,\ldots k-2)} \prod_{\alpha=1}^{k-1}\delta^{4|4}(t_{\alpha a}{\cal Z}^D_{a})\cr
&= \int \prod_{a=1}^n \frac{dw_a}{w_a} e^{iw_az_a} \delta^{3|4}(w_a\tilde Z_a){\cal Y}_{n,k-1}({\cal Z}^D)\,.
\end{align}
From this expression, it is clear that the process can be iterated until we reduce $k$ all the way to unity. 

\section{N$^2$MHV residues}

Let us look at the simplest non-trivial example, $k=2$. Our general result simplifies to
\be\label{newRnk}
{\cal Y}_{n,2}({\cal Z}) =  \int \prod_{a=1}^n \frac{dw_a}{w_a} e^{iw_az_a} \delta^{3|4}(w_a\tilde Z_a) \int \frac{1}{\text{Vol}(GL(1))}\prod_{a=1}^{n} \frac{dt_a}{t_a}\delta^{4|4}(t_a {\cal Z}^D_a) \,.
\ee
There are several comments we would like to make about this formula.
 
First, note that unlike the original Grassmannian integral \eqref{Grassmannian}, we now have $2n$ factors in the denominator. As in  \eqref{d}, we have to impose $d=2n-12$ constraints in order to compute a residue. However note that we cannot pick any $d$ out of the $2n$ factors and expect to find a rational function of $Z_a=(\tilde Z_a, z_a)\,$. Most choices lead to what we would like to call ``singular residues", which are distributions in the $Z$'s and therefore impose additional constraints on the external momenta. In what follows, we will develop a systematic procedure to identify the regular residues of \eqref{newRnk}. 

There is, however, an important caveat. The counting that gave us the number of conditions to be imposed in order to obtain a residue was derived from the original expression \eqref{oldRnk}. From there, it is clear that we have $k(n-k)$ variables and $4k$ delta functions; it follows that we need to impose $k(n-k-4)$ conditions to localize the integral and obtain a residue. Let us do a similar counting in the derived expression \eqref{newRnk}. A naive counting gives us $2n-1$ variables and seven delta functions, which would suggest imposing $2n-8$ conditions, different from the counting we know to be correct. It is therefore unclear, from the point of view of \eqref{newRnk}, why imposing $2n-12$ conditions should lead to a rational function of the $Z$'s, or whether it even makes sense to talk about residue theorems in this new formalism.

The situation is very similar to the issue of defining residues in the original twistor variables. There too, there is a naive mismatch between the number of conditions and integration variables
\footnote{We would like to thank Freddy Cachazo for pointing this out to us.}. 
In both these cases, it would be desirable to have a mathematical understanding of why such a definition leads to well-defined residues and consequently, of residue theorems and we do not have this understanding at present. In both cases, one appeals to the formulation in momentum space (related via Fourier transform) in which the counting is transparent. 

What we will show is that it is possible to mirror the known residue theorems involving D-minors by identifying a specific combination of $w$'s and $t$'s with a given consecutive minor $D(a,a+1)$. This will guide us in writing out residue theorems in the new variables and we will recover not only the simplest residue theorems, but also the supposedly more complicated ones involving composite residues, in a natural manner. 

The main advantage of this approach is that typically there are many configurations of the $(t_a, w_a)$ variables that correspond to the vanishing of a given set of $d$ minors. For $8\le n\le12$ we find that the number of solutions always agrees with the Littlewood-Richardson coefficient that was quoted in \cite{ACCK0907}. We interpret this as evidence that the new variables do in some sense break down each minor into its irreducible components. As we will see later, a consequence of this fact is that the composite residues and the non-composite ones are equally accessible in this formalism. 

\subsection{Vanishing minors in new variables}\label{map}

In this section we show how the vanishing of either of the  $(t_a, w_a)$ variables relates to the vanishing of specific $D$-minors. Recall that in our gauge the matrix $D$ takes the following form:
\be
D = \left(\begin{array}{ccccc}
w_1 & w_2 &\ldots & & w_n \\
D_{21}& D_{22}& \ldots & & D_{2n}
\end{array}\right) \,.
\ee
The $D_{2a}$ are related to the $t_a$ variables as follows:
\be\label{defta}
t_a = \frac{w_{a+1} D_{2,a}-w_{a}D_{2,{a+1}}}{w_a^2w_{a+1}}\ .
\ee
However, only $n-1$ of these are linearly independent, so we need to use $n-1$ of these equations along with the linearly independent equation
\be
\sum_{a}D_{2a}w_a = 0 \,.
\ee
From these we can solve for the $D_{2a}$ and take the limit of these functions as various $t$'s and $w$'s are set to zero. We find the following interesting cases:
\begin{itemize}
\item $t_{a}=0$ and $w_a \ne 0\,$: The columns $a$ and $a+1$ are proportional and $D(a,a+1)=0\,$, with no other minor being set to zero. Conversely, it is always possible to solve $D(a,a+1)=0$ by setting $t_a=0\,$. 
\be
t_{a}=0 \leftrightarrow D(a,a+1)=0
\ee
\item $w_a = 0$: Solving the equations for the $D_{2b}$, we find that this leads to $D_{2a}=0$. The matrix $D$ therefore takes the form
\be
D = \left(\begin{array}{cccccccc}
w_1 & \ldots &w_{a-1}   &0 & w_{a+1}  &\ldots & & w_n \\
D_{21} &\ldots& D_{2,a-1} &0& D_{2,a+1}&\ldots & &D_{2n}
\end{array}\right) \,.
\ee
In terms of consecutive $D$-minors, this is equivalent to $D(a-1,a)=D(a,a+1)=0$. These are two conditions, but one observes that when we set $w_a=0$ the variable $t_a$ drops out of the system of equations, so we are free to impose an additional condition on $t_a$ at no extra cost. 
In the analysis that will follow we will only consider contours such that the pole $w_a=0$ can be picked up only on the locus where $t_a=0\,$. We will restate our earlier conclusion as follows:
\begin{align}
\{t_a=w_a = 0\} 
&\leftrightarrow 
D=\left(\begin{array}{cccccccc}
w_1   & w_2    &\ldots  & 0&w_{a+1}& &\ldots & w_n \cr
D_{21}& D_{22}& \ldots & 0& D_{2,a+1}             & &\ldots & D_{2n} 
\end{array}\right) \cr
&\leftrightarrow D(a-1,a)=D(a,a+1)=0 \,.
\end{align}
Setting more pairs of $(w_a, t_a)$ just sets more columns to zero. However if two adjacent columns are removed, $w_a=t_a=w_{a+1}=t_{a+1}=0$, this gives a composite residue. One way to understand this is to notice that such a configuration cannot be specified in terms of minors involving only consecutive columns. For instance, a minimal gauge invariant characterization of this solution could be given by the set of equations $D(a-1,a)=D(a,a+1)=D(a+1,a+2)=D(a-1,a+1)=0\,$. 

\item There is one more way to obtain a composite residue:
\begin{align}\label{composite}
\{t_a=w_a =t_{a-1}= 0\} &\leftrightarrow \left(\begin{array}{ccccccccc}
w_1   & w_2    &\ldots  & w_{a-1}& 0&w_{a+1}&\ldots & w_n \cr
D_{21}& D_{22}& \ldots & x\, w_{a-1}            & 0& x\, w_{a+1}            &\ldots & D_{2n}
\end{array}\right)\cr
&\leftrightarrow D(a-1,a)=D(a,a+1)=D(a-1,a+1)=0 \,.
\end{align}
As in the case of two consecutive vanishing columns, we cannot express the constraint exclusively in terms of minors that involve consecutive columns of $D\,$, and this is a sign of the fact that this configuration corresponds to a composite residue. 
\end{itemize}

\subsubsection*{Vanishing residues}

We will not evaluate the residues explicitly. However, since our goal is to classify the residues, it is useful to know when certain configurations lead to vanishing residues. From explicit calculations \cite{ABC1008}, it has been shown that if it is possible to choose a gauge in which more than $n-5$  entries in a row are zero, then the residue vanishes. This is most easily seen in the calculation of residues using momentum twistors. There are $4k$ bosonic delta functions in \eqref{oldRnk}. Let us look at these as a set of 4 delta functions for each row of the $D$. Then we see that if we consider a configuration of $D$ with more than $n-5$ zeroes in a row, it becomes impossible to solve for the delta functions without constraining the momentum twistors. This leads to a vanishing residue.

We will now analyze in detail the $n=8$ example, which is the simplest nontrivial case, and then we will present our results for $n\leq12$. The goal is to obtain a classification of the non-vanishing residues and of the relations between them, and to identify a minimal set of independent residues from these relations. 

\section{$n=8$: A case study}

In order to check the results of the previous section, let us apply the rules we have found to the case of 8 particles and count the number of non-trivial residues. $d=4$ in this case and we need to set four minors to zero. Without including the composite ones, we expect to find $140$ residues: there are ${8\choose 4}=70$ ways to choose four minors out of eight and for each such choice there are two solutions \cite{ACCK0907}. 

\subsection{Counting residues}\label{residuecount}

In the $(w,t)$ variables, the counting is rather simple for the $n=8$ case. There are only three non-trivial cases to consider:
\begin{itemize}
\item $w_a = w_b=t_a=t_b=0$: This corresponds to ${8\choose 2}=28$ residues. However, for $b=a\pm 1$ these are composite residues, so $8$ out of $28$ residues are composite. 
\item $w_a=t_a=t_b=t_c=0$: Before counting these residues, note that if $b$ and $c$ are consecutive the residue will vanish. The reason is that we can gauge fix column $b$ to 
$\bigl( \begin{smallmatrix} 1\\ 0
\end{smallmatrix} \bigr)$
and then the condition $t_b=t_{b+1}=0$ will force the columns $b\,$, $b+1$ and $b+2$ to be proportional, resulting in three consecutive zeroes. Considering that column $a$ is also set to zero, this residue must vanish, as discussed at the end of section \ref{map}.  \newline
Keeping this in mind, we get a total of $8\times 14= 112$ non-vanishing residues. Whenever either $b$ or $c$ equals $a-1$ the residue is composite, so there are $32$ composite residues and 80 non-composite residues of this type. 
\item $t_a=t_b=t_c=t_d=0$: Proceeding as in the previous case, it is easy to see that whenever three or more of the $t$'s are consecutive the residue will vanish. This gives $38$ non-trivial residues, all non-composite.
\end{itemize}

As we mentioned above, we expect two solutions for each set of four vanishing minors. This is precisely what we see directly from the $(w,t)$ variables, with two exceptions:
\begin{align}
D(12)=D(34)=D(56)=D(78)&=0\cr 
\text{and}\quad D(23)=D(45)=D(67)=D(81)&=0\,.
\end{align}
In these cases we see only one solution: $t_1=t_3=t_5=t_7=0$ and $t_2=t_4=t_6=t_8=0$ respectively. These correspond to the leading singularities of the 4-mass box  diagrams which have also been discussed in \cite{ACCK0907}. Even in our new variables, we cannot split these pairs of residues and so these two conditions define four residues. 

Accounting for the two irreducible solutions, we find a total of $20+80+38+2=140$ non-composite residues, exactly what was found in \cite{ACCK0907}. In addition, we find $8+32=40$ composite residues.

\subsection{Global Residue Theorems for $n=8$}

The residues we counted in the earlier section are not all independent. They satisfy relations such us
\be\label{residueD}
D(12)D(23)D(56) \big[D(34) + D(45) + D(67) + D(78) + D(81) \big] = 0 \,,
\ee
where with a standard abuse of notation we are denoting residues by the corresponding vanishing minors.

It is useful to understand the relations between residues as descending from a global residue theorem. We will follow the discussion in \cite{ACCK0907} and refer to that paper as well as the mathematical literature \cite{GH, Tsikh, Yuzhakov} for details. Consider the $M$-form 
\be\label{resform}
\omega=\frac{h(z_i)dz_1\wedge \ldots \wedge dz_M}{f_1(z_i)\ldots f_M(z_i)}\,.
\ee
Let $F_i=\{z\in \mathbb{C}^M:f_i(z)=0\}$ be the $M-1$ dimensional subspace associated with $f_i$ and let $Z$ be the intersection of all such hypersurfaces. Here $Z$ is assumed to be a discrete set of points. Then, the global residue of $h$ with respect to the map $f$ is defined as
\be\label{theo}
\text{Res}_f(h) = \sum_{a\in Z} \text{res}(\omega)_a \,.
\ee
The global residue theorem states that:
 \be
\text{if}\qquad deg(h)<deg(f_1)+\ldots + deg(f_M)-M\qquad \text{then}\qquad \text{Res}_f(h)=0\ .
 \ee 

One can check that the relation \eqref{residueD} can be derived, in the original Grassmannian formulation, by applying the global residue theorem \eqref{theo} to the following set of $f_i$'s: 
\begin{align}
f_1 &= D(12)\cr
f_2&= D(23)\cr
f_3&=D(56)\cr
f_4&=D(34)D(45)D(67)D(78)D(81)
\end{align}
Note that, in all our applications, the degree $M$ of the form \eqref{resform} coincides with the number of conditions $d$ derived in \eqref{d}. So for $n=8\,$, we need $M=4$ functions $f_i\,$.

We can use the explicit map of Section \ref{residuecount} to rewrite the terms of the relation \eqref{residueD} as
\begin{align} \label{twoforone}
D(12)=D(23)=D(56)=D(78)=0&\leftrightarrow \left\{\begin{array}{c} t_1=t_2=t_5=t_7=0\\ w_2=t_2=t_5=t_7=0\end{array}\right. \cr
D(12)=D(23)=D(56)=D(34)=0&\leftrightarrow\left\{\begin{array}{c}  t_1=w_3=t_5=t_3=0\\ w_2=t_2=t_5=t_3=0\end{array}\right. \cr
D(12)=D(23)=D(56)=D(34)=0&\leftrightarrow \left\{\begin{array}{c}  t_1=t_2=t_5=t_4=0\\ w_2=t_2=t_5=w_5=0\end{array}\right. \cr 
D(12)=D(23)=D(56)=D(67)=0&\leftrightarrow  \left\{\begin{array}{c}  t_1=t_2=t_5=t_6=0\\ w_2=t_2=w_6=t_6=0\end{array}\right. \cr 
D(12)=D(23)=D(56)=D(81)=0 &\leftrightarrow \left\{\begin{array}{c} t_1=t_2=t_5=w_1=0\\ w_2=t_2=t_5=t_8=0 \end{array}\right. \,.
\end{align}
Notice that we see here the two explicit solutions to the vanishing of four minors in the $(w,t)$ variables. In order to write down the corresponding residue theorem in terms of the $(w,t)$ variables, we will heuristically identify the combination $w_{a+1}t_{a}$ with the D-minor $D(a,a+1)$. The main motivation for this comes from the fact that whenever the minor $D(a,a+1)$ is set to zero in a non-composite residue, these factors are never set to zero simultaneously - as you can check for instance in \eqref{twoforone} above. The two factors are however both set to zero in a composite residue, as in \eqref{composite}\,. This suggests that the $w_{a+1}$ and $t_a$ factors should be roughly thought of as two pieces of the $(a,a+1)$ minor.  

Following this logic, let us choose the four functions in \eqref{resform} to be
\begin{align}\label{residuewt}
f_1 &= (w_2 t_1)   \cr
f_2 &= (w_3 t_2) \cr
f_3 &= (w_6 t_5) \cr
f_4 &= (w_8 t_7) (w_4 t_3) (w_5 t_4) (w_7 t_6) (w_1 t_8) \,.
\end{align}
Keeping only the non-vanishing residues, we get
\begin{multline}
\{t_1, t_2, t_5, t_7\}+ \{t_1, w_3, t_5, t_3\} + \{t_1, t_2, t_4, t_5\}+\{t_1, t_2,  t_5, t_6\}+\{t_1, t_2, t_5, w_1\} \cr
+\{w_2, t_2, t_5, t_7\} + \{w_2, t_2, t_5, t_3\}+\{w_2, t_2, w_5, t_5\}+\{w_2, t_2, w_6, t_6\}+\{w_2, t_2, t_5, t_8\}=0\,,
\end{multline}
which is what we expected from the map  \eqref{twoforone}. 

\subsection{Counting independent residues}

We can deduce the number of independent residues by analyzing systematically all possible residue theorems. For $n=8$ there are five different kinds of residue theorems, associated with different partitions of $8\,$ into four pieces:

\begin{itemize}

\item $1+1+1+5$: These are theorems of the kind \eqref{residuewt}, where we set $f_1\,$, $f_2$ and $f_3$ equal to one minor each and set $f_4$ equal to the product of the remaining five minors. There are ${8\choose 3}=56$ residues theorems of this type. 

\item $1+1+2+4$: In this case we choose two $f$'s to be single factors, one $f$ to be the product of two minors and the last $f$ to be the product of the remaining four minors. There are  ${8\choose 2}\times {6\choose 2}=420$ theorems of this type. 

\item $1+1+3+3$: There are ${8\choose 2}\times {6\choose 3}=560$ theorems in which two of the $f$'s have one minor each while the remaining two have three each.  

\item $1+2+2+3$: There are $8\times {7\choose 2}\times {5\choose 2} = 1680$ such residue theorems. 

\item $2+2+2+2$: There are ${8\choose 2}\times {6\choose 2} \times {4\choose 2}=2520$ such residue theorems. 

\end{itemize}
This counting tells us that there are overall $5236$ residue theorems involving the $138$ non-composite residues we described in section \ref{residuecount}. Using a simple algorithm implemented in Mathematica, we have generated a $5236\times 138 $ matrix that for each row has elements $1$ or $0$ depending on whether a specific residue appears in a give theorem or not. Mathematica could easily compute the rank of this matrix, which turned out to be equal to $70$. So, including the two extra residues which cannot be split in our variables, we find that there are $72$ independent residues for the $n=8$, $k=2$ case. All other residues can be obtained from these ones by using the residue theorems. 

\subsubsection{Residue theorems involving composite residues}

For completeness, we would like to describe how to write down in the new variables the residue theorems with composite residues. These theorems necessarily involve splitting off a given combination $(w_{a+1}t_a)$ and assigning each factor to different functions $f_i$. However, we cannot split the factors arbitrarily. Lacking a full understanding of how to derive the theorems directly from \eqref{newRnk}, we will try and derive the rules more heuristically.

The basic idea, for which we refer to \cite{ACCK0907}, is that on the locus where the minor $D(a,a+1)$ vanishes, the minor $D(a+1,a+2)$ factorizes. This can be easily checked from the Pl\"ucker relation
$$
D(a+1\, a+2)D(a \,c ) = D(a \,a +1)D(a+2 c)+D(a+2 \,a)D(a+1\, c) \,, 
$$
(where we assume that $D(a \,c )$ is nonzero).
We don't expect to be able to break down a minor into two factors unless we are on the locus where one of its adjacent minors vanish. Rephrasing this idea, we don't expect to be able to split the combination $(w_{a+1}t_a)$ unless the successive minor $(w_{a+2}t_{a+1})$ vanishes in \emph{all} terms of the residue theorem.  In addition, for consistency, in the new configuration obtained by splitting $w_{a+1}$ and $t_a$, the composite residue must be non-zero. These simple rules eliminate a lot of inconsistent residue theorems and match our expectations in all the cases we have considered. 

Let us discuss one example. 
Consider the set of functions \eqref{residuewt}. Because either $w_3$ or $t_2$ appear in each term of the residue theorem, according to the rules proposed above we are allowed to split $(w_2t_1)$ as follows: 
\begin{align}\label{compositewt}
f_1 &= t_1 Ê\cr
f_2 &= (w_3 t_2) \cr
f_3 &= (w_6 t_5) \cr
f_4 &= (w_8 t_7) (w_4 t_3) (w_5 t_4) (w_7 t_6) (w_1 t_8) w_2 \,.
\end{align}
Note that we couldn't have split, say, $(w_6t_5)\,$, because the minor $ (w_7 t_6)$ is not common to all the terms. The residue theorem that follows from \eqref{compositewt} is
\be
\{t_1\,t_2\, t_5\, t_7 \} + \{t_1\, t_3\, t_5\, w_3 \} + \{t_1\, t_2\, t_4\, t_5 \} + \{t_1\, t_2\, t_5\, t_6 \} + \{t_1\, t_2\, t_5\, w_1 \} + \{t_1\, t_2\,t_5\, w_2 \} = 0\,.
\ee
The first five terms in this equation also appear in the original residue theorem \eqref{residuewt}, while the last one is the composite residue. 

Similarly, it is possible to use the $(w_a, t_a)$ variables to express both the usual residue theorems and those that include composite residues in a unified and natural manner. Of course, since every composite residue is just a linear combination of non-composites, this does not change the counting of independent residues that we performed earlier.

\section{Higher $n$}

Most of the discussion for $n=8$ particles can be directly carried over to the higher $n$ case, as long as we restrict to $n\le 12\,$, so we will be brief in our discussion and merely list the results for the higher values of $n\,$.

\subsection{$n=9$}

 We have checked that in this case too it is possible to see explicitly the distinct solutions that follow from setting $d=2n-12$ minors to zero. Let us illustrate how this works with an example. For $n=9\,$, we need to impose $d=6$ conditions to define a residue. Consider the set of conditions 
\be
D(12)=D(23)=D(34)=D(56)=D(78)=D(89)=0 \,.
\ee
They can be  rewritten  as
\be
t_1 w_2=t_2 w_3=t_3 w_4=t_5 w_6=t_7 w_8=t_8 w_9 = 0\,
\ee
and one can check that the non-vanishing solutions are the residues
\begin{multline}
\{t_1, t_2, t_3, t_5, t_7, t_8\}\,, \{w_2, t_2, t_3, t_5, t_7, t_8\}\,, \{t_1, w_3, t_3, t_5, t_7, t_8\}\,,\cr
\{w_2, t_2, t_3, t_5, w_8, t_8\}\quad\text{and}\quad \{t_1, w_3, t_3, t_5,w_8, t_8 \}\,. 
\end{multline}
There are five solutions, as expected on general grounds \cite{ACCK0907}.

We find the same number of solutions in almost all cases, with precisely nine exceptions. These are all of the form $\{w_a, t_a, t_{a+1}, t_{a+3}, t_{a+5}, t_{a+7}\}$ and they are interesting because they arise from 8-particle residues $\{t_{a+1}, t_{a+3}, t_{a+5}, t_{a+7}\}$ via an inverse soft-factor, as described in \cite{ABC1008}. These particular $8$-particle residues were already discussed in Section \ref{residuecount}, where we pointed out that each of these corresponds to two residues.

So there are a total of ${9\choose 6}\times 5 = 420$ non-composite residues, out of which we can distinguish $411$ in our new variables. It only remains to classify these by writing out the residue theorems and finding out which of the residues are independent. There are only three types of residue theorems to consider, corresponding to the three partitions of $9$ into six parts:
\begin{itemize}
\item $1+1+1+1+1+4$ \qquad $\longrightarrow$\qquad ${9\choose 5}=126$ theorems 
\item $1+1+1+1+2+3$ \qquad $\longrightarrow$\qquad ${9\choose 4}\times {5\choose 2}=1260$ theorems 
\item $1+1+1+2+2+2$ \qquad $\longrightarrow$\qquad ${9\choose 3}\times {6\choose 2}\times {4\choose 2}=7560$ theorems\ .
\end{itemize}
Computing the rank of the matrix of residue theorems, as discussed earlier for $n=8$, we find that there are $84$ independent residues written in the $(t,w)$ variables. Including the $9$ that cannot be split, we find a total of $93$ independent non-composite residues. 

\subsection{$n\le 12$}

As long as $n< 12\,$, it is possible to follow the reasoning we have outlined in the previous sections and find the non-composite residues, use the residue theorems to find the linearly independent residues and write the composites as linear combinations of this basic set. We list below a few of the interesting results for these cases.

\begin{itemize}

\item For higher $n\,$, there is only one subtlety to consider when counting the residues. This arises when the $8$-particle residues which correspond to a $4$-mass box are lifted to a higher dimensional residue via inverse soft factors. In such cases, we saw that even in the $(w,t)$ variables, we could not split the two residues that solved the vanishing of the four minors. So even when lifted to a higher $n$ residue, each such configuration should count as two residues. 

\item We have already seen that for $n=8$ and $n=9$ we find, respectively, $2$ and $5$ solutions for each set of $d$ vanishing minors. For $n=10\,$, solving for  $d=8$ vanishing minors, we find $14$ solutions. For $n=11$ and $d=10$ we find $42$ solutions. These distinct solutions can be explicitly described by the vanishing of specific combinations of $w$'s and $t$'s.

\item For $n=12$ there is only one set of $d=12$ consecutive minors one can write. Setting them to zero, we find $130$ solutions in the $(w_a, t_a)$ variables. Three of these residues are obtained from the four-mass box eight particle residue, so there would seem to be $133$ solutions. There is however precisely one residue theorem that makes one of them linearly dependent on the others, yielding $132$ independent (non-composite) residues. 

\item All of these numbers are consistent with the general result quoted in \cite{ACCK0907}, which is that the number of solutions expected to setting $d=2n-12$ minors to zero is
\be\label{LRcoeff}
\text{\#}=\frac{(2n-12)!}{(n-6)!(n-5)!}\ .
\ee
Being able to reproduce this number for all the cases we have considered is a good consistency check of our methods. 

\item Listing the residues in the $(w,t)$ variables, it becomes clear that already for $n >10$ there are no more ``new" residues, in the sense that each non-vanishing residue has to have at least $n-10$ columns set to zero. Translated in our variables, this means that for $n>10$ every residue will result from imposing at least  enough $w_a=t_a=0$ conditions to reduce the effective size of the matrix to $n=10\,$.

\end{itemize}

\subsection{$n>12$}
For $n\ge 12$, our approach has to be modified. This is because the number of conditions $d$ becomes larger than the number of consecutive minors. Although listing  the residues seems straightforward enough, what is not clear to us is the role played, if at all, by residue theorems for large $n$. So a classification of independent residues remains an outstanding problem.
Deriving the Littlewood-Richardson coefficient in \eqref{LRcoeff} for $n> 12$ is also an open problem, that might require finding a description of the non-consecutive minors in terms of the $(w_a, t_a)$ variables. We hope to address these issues in the future. 

\section{Discussion and Summary of Results}

In this work we have put forward a proposal for the classification of residues of the Grassmannian integral and we have listed all  residues for $k=2$ and $n\le 12$. By suitably adapting the procedure that takes the spinor-helicity variables to momentum-twistors, we introduced new variables that factor the minors that appear in the integral into irreducible components. By irreducible we mean that these variables are the simplest building blocks in terms of which any residue can be specified.

This has a direct consequence for the counting of solutions to a given a set of vanishing minors. 
The number of solutions one expects can be obtained from the Littlewood-Richardson decomposition of the self-intersection of a Schubert cycle; this has been checked for many examples in \cite{ACCK0907}. For instance, for $k=2\,$, one expects two solutions for $n=8$, five solutions for $n=9$ and so on, up to $132$ solutions for $n=12$. The fact that these distinct solutions can be explicitly written out in the new variables is an important check on the validity of our approach. 

Another attractive feature is how both composite and non-composite residues appear on the same footing: both of these correspond to the vanishing of some combination of $w$'s and $t$'s and the composite residues are no more difficult to handle than the non-composite ones. For small enough values of $n$, we used the residue theorems in the new variables involving only non-composite residues to completely classify all the independent residues. The computation is quite simple and it can be easily implemented in Mathematica.

It will, of course, be challenging and interesting to extend our analysis to higher values of $n$, but especially to higher values of $k$ where there are not many results available in the literature. The key would be to identify the combination of new variables that corresponds to the $k\times k$ minor in the general case. Once the identification is made, and the map between the vanishing minors and the new variables found, the analysis should go along the lines discussed for $k=2$ in this work. 

However there are many open questions still to be addressed. It would be interesting to see if the iterative formula can be useful in the evaluation of the residues. Also, as we discussed already in the main text, the status of residue theorems in the new variables needs to be better understood, given that the number of delta functions seems to be insufficient to localize the integral. Similarly, the precise nature of the Fourier transform and the role played by the $3$-momentum conserving delta functions are some points to be clarified. We hope to address these issues in the future. 

\section*{Acknowledgements}

We are extremely grateful to Freddy Cachazo for suggesting the problem to us and for the many helpful suggestions and useful discussions over the past months. We also thank him for his comments on the draft. We would like to thank David Skinner for many helpful discussions. We would also like to thank Nima Arkani-Hamed and Freddy Cachazo for their insightful lectures during the mini-course ``Space-time, Quantum Mechanics and Scattering Amplitudes", organized at the Perimeter Institute in August, 2010. The work of ED is supported by the Perimeter Institute for Theoretical Physics. Research at the Perimeter Institute is supported by the Government of Canada through Industry Canada and by the Province of Ontario through the Ministry of Research and Innovation.

\end{document}